\documentclass[12pt]{article}

 \newread\testifexists
 \def\GetIfExists #1 {\immediate\openin\testifexists=#1
     \ifeof\testifexists\immediate\closein\testifexists\else
     \immediate\closein\testifexists\input #1\fi}

\usepackage{amsfonts}
 \usepackage{amssymb}
 \usepackage{graphicx} \usepackage{epstopdf}
 \immediate\openin\testifexists=e
     \ifeof\testifexists\immediate\closein\testifexists\else
     \immediate\closein\testifexists\input e\fipsf

 \def\Bbb#1{\setbox0=\hbox{$\tt #1$}  \copy0\kern-\wd0\kern .1em\copy0}

 \def\bbf#1{\setbox0=\hbox{$#1$} \kern-.025em\copy0\kern-\wd0
         \kern.05em\copy0\kern-\wd0 \kern-.025em\raise.0433em\box0}

 \immediate\openin\testifexists=a
     \ifeof\testifexists\immediate\closein\testifexists\else
     \immediate\closein\testifexists\input a\fimssym.def  

                     \newcommand{\fn}{\footnote}
              \newcommand{\nm}{\nonumber}
 \newcommand{\be}{\begin{eqnarray}}             \newcommand{\ee}{\end{eqnarray}}
 \newcommand{\bi}[1]{\begin{itemize}\item[#1]}           \newcommand{\ei}{\end{itemize}}
 \newcommand{\eqn}[1]{(\ref{#1})}

 \newcommand{\newsec}[1]{\setcounter{equation}{0}}

 \newcommand{\crlb}[1]{\label{#1}\\[2pt]}
 \newcommand{\eela}[1]{\quad\hbox{\scriptsize{#1}}\label{#1}\end{eqnarray}}
 \newcommand{\eelb}[1]{\label{#1}\end{eqnarray}}
 
 \newcommand{\newsecb}[2]{\section{#1}\label{#2}\setcounter{equation}{0}}

 \newcommand{\nolabels} {\def\eel{\eelb} \def\crl{\crlb} \def\newsecl{\newsecb}\def\bibiteml{\bibitem}\def\citel{\cite}}

\newcommand\publishversion{\nolabels\setlength{\textheight}{9.1in}\setlength{\oddsidemargin}{0in}
    \setlength{\textwidth}{6.3in}\setlength{\topmargin}{-0.5in}}

          \def\g{\gamma}      
          
             \def\m{\mu}
 \def\f{\phi}                \def\n{\nu}

       \def\W{\Omega}

  \def\ra{\rightarrow}
  
 \def\dd{{\rm d}}  \def\bra{\langle}   \def\ket{\rangle}

 \def\ffract#1#2{\raise .2 em\hbox{$\scriptstyle#1$}\kern-.3em/
                 \kern-.2em\lower .15 em \hbox{$\scriptstyle#2$}}

\def\bmatrix{\begin{matrix}} \def\ematrix{\end{matrix}} \def\bpmatrix{\begin{pmatrix}}\def\epmatrix{\end{pmatrix}}
\def\bcenter{\begin{center}} \def\ecenter{\end{center}}

\def\lowerheightfig#1#2#3{\(\raise-#1\hbox{\includegraphics[height=#2]{#3}}\)}
\def\lowerwidthfig#1#2#3{\(\raise-#1\hbox{\includegraphics[width=#2]{#3}}\)}

\publishversion
\begin{document} 
 \begin{titlepage}

\title{\normalsize \hfill ITP-UU-14/25  \\ \hfill SPIN-14/19
\vskip 20mm \Large\bf Local Conformal Symmetry: the Missing Symmetry Component for Space and Time}

\author{Gerard 't~Hooft}
\date{\normalsize Institute for Theoretical Physics \\
Utrecht University \\ and
\medskip \\ Spinoza Institute \\ Postbox 80.195 \\ 3508 TD Utrecht, the Netherlands \smallskip \\
e-mail: \tt g.thooft@uu.nl \\ internet: \tt
http://www.staff.science.uu.nl/\~{}hooft101/}

\maketitle

\begin{quotation} \noindent {\large\bf Abstract } \medskip \\
Local conformal symmetry is usually considered to be an approximate symmetry of nature, which is explicitly and badly broken. Arguments are brought forward here why it has to be turned into an exact symmetry that is \emph{spontaneously} broken. As in the B.E.H. mechanism in Yang-Mills theories, we then will have a formalism for disclosing the small-distance structure of the gravitational force. The symmetry could be as fundamental as Lorentz invariance, and guide us towards a complete understanding of physics at the Planck scale.
\end{quotation}

  \vfill \flushleft{Essay written for the Gravity Research Foundation 2015 Awards for Essays on Gravitation.\\ Submitted  \today}
 \end{titlepage}
 \eject
 \setcounter{page}{2}
All important physical systems have a built-in scale in them, and for that reason, conformal transformations may appear to be useless as a symmetry group; at best, this symmetry is badly broken. The topic of this essay, however, is that one could try to reason differently. 

Why have symmetries always been so instrumental for understanding nature? The answer is this: if we know the laws of nature in one particular domain, the laws in other domains can be obtained by applying a symmetry transformation. For instance, translation symmetry tells us that the laws are the same ones everywhere in space and in time. Rotational symmetry tell us that they are the same in all directions. Furthermore, Lorentz transformations, replacing the Galile\"\i\ transformations,  tell us how a moving particle behaves if we know what that particle does when at rest. In theories without Lorentz invariance, moving particles are altogether different from stationary ones.

Why then is physics still so difficult? Well, we still do not know what happens at higher energies even if we do understand the laws at low energies, or more to the point: small time and distance scales seem not to be related to large time and distance scales. Now, we argue, this is because we fail to understand the symmetry of the scale transformations. This symmetry, of which the local form will be local conformal symmetry, \emph{if exact}, should fulfil our needs. Since the world \emph{appears} not to be scale invariant, this symmetry, if it exists, must be spontaneously broken. This means that the symmetry must be associated with further field transformations, leaving the vacuum not invariant. It is this implementation of the symmetry that we should attempt to unravel from the evidence we have.

We are dealing with a component in the space-time symmetry group (the Poincar\'e group) that both Lorentz and Einstein may have seen, but dismissed. Lorentz derived the invariance group named after him as a property of electro-magnetism alone. Now \emph{that} system, described by Maxwell's equations, does have conformal symmetry; the transformations \(x^\m\ra\,x^\m\,/\,x^2\)  are to be added to the other generators of the Poincar\'e group. What this means is that, if we only use light rays to measure things,  absolute sizes and time spans cannot be observed, only relative ones.\fn{The reader might be worried about invoking conformal symmetry: did that not transform straight lines into circles, and shouldn't we be able to distinguish straight lines from circles using light? The answer is yes, but here we use \emph{four} dimensional conformal symmetry, which does transform light cones into light cones.}

Suppose that Einstein started from here to formulate his special theory of relativity. Can one compare observers with scaled observers? Should we not have ended up with the conformal extension of the Poincar\'e group? Let us have a look at the elementary principles of relativity. Now however, we decide \emph{only} to use light rays for measuring things, and suppose we would wish to set up a theory for gravity. How would that work? 

To be sure, we promise to put the scales of things back in the world of physics in the end, by saying that conformal symmetry will be \emph{spontaneously} broken, but we haven't reached that point yet; we first wish to describe the world in the symmetric picture. This is a picture of the world that may be well hidden from our eyes today, in the same way that exact local \(SU(2)\) symmetry in the weak interactions has long been hidden from us, before the BEH mechanism was understood.

Considering now conformal gravity\cite{PM}, imagine an observer who neither has a ruler, nor a clock, just light rays. Einstein could not have started with the elementary line element \(\dd s\), but he would have to consider only the light-like geodesics, described by
	\be \dd s^2=g_{\m\n}\dd x^\m\dd x^\n=0\ . \eel{lightelement}
This means that, at every space-time point \(x\) separately, all relative values of the metric tensor components \(g_{\m\n}(x)\) could be used, but not the common factor, so we would have to consider a `pseudo' tensor \(g_{\m\n}\) that is defined apart from this factor:
	\be g_{\m\n}(\vec x,t)\ \ra \ \W^2(\vec x,t)g_{\m\n}(\vec x,t)\ , \eel{confmetric}
an equation to be regarded as a one-dimensional \emph{local} gauge transformation.\fn{Sometimes this is referred to as the Weyl group, but Weyl introduced the associated vector field, which we avoid here.}

If we look at just two of the four coordinates \(x^\m\), light rays are invariant under
	\be \matrix{x^+\ =\ x+c\,t& \ra& \g_1x^+\ ,\nm\\ x^-\ =\ x-c\,t& \ra& \g_2x^-\ .} \eel{confLorentz}
Normally, one puts \(\g_1\g_2=1\) (the Lorentz boost), but in the case of conformal symmetry, one simply drops this constraint. The importance of this could be that it enables us to reach the small distance limit, and with that the high energy domain, in the coordinates \(x^+\) and \(x^-\) separately.

Now consider the scalar component \(R\) of the Ricci curvature. If we subject the metric tensor to the transformation \eqn{confmetric}, it transforms as
	\be R\ \ra\ \W^{-2}R-6\W^{-3}D^2\W\ , \eel{Rtransf}
where \(D_\m\) is the covariant derivative, and this means that we can always choose \(\W\) in such a way that \(R\) is set equal to zero everywhere.  According to the Einstein equations, the trace of the energy-momentum tensor is proportional to the scalar Ricci curvature. This means that the trace of the energy-momentum tensor can also be transformed away. To be precise: \emph{by using light rays alone, one cannot detect the scalar component of the energy-momentum tensor.} It is ill-defined.\fn{Not to be confused with the property of conformally invariant matter that has this trace vanish identically; space limitations keep me from explaining this situation adequately.}  But if you put it equal to zero, then \(\W\) is fixed up to boundary terms, and then the 9 remaining components of the Ricci curvature and the energy-momentum tensor are fixed (up to boundary effects).

By adding local conformal transformations as a fundamental gauge group, space-time curvature simplifies somewhat: the usual Riemann curvature tensor with its 20 independent components is replaced by the Weyl tensor, which has only 10.

The \emph{vacuum state} would normally have \(R=0\), and now we see that this is not invariant. This is why we say that the vacuum breaks local conformal symmetry `spontaneously' (as in the BEH mechanism). At first sight, it may seem that the vacuum does not break \emph{global scale} invariance. However, since Nature is clearly not invariant under scale transformations, we must assume that, under scale transformations, the vacuum transforms into another, as yet unknown state. We claim this to be of crucial importance to understand the properties of black holes.

In a nut shell, the situation with black holes is this: an observer falling into a black hole, does not notice any substantial changes in the black hole's mass as he or she passes the horizon, but an outside observer may see the mass of a black hole graudually shrink to zero due to the emission of Hawking radiation. To reconcile their findings, one may now assume that the two observers disagree about the scale of things. What is a local vacuum for one observer, is something else for the other. Of course it is. One observer sees Hawking particles as matter, the other as part of his vacuum. The author believes that this will be the beginning of a real resolution of the black hole information paradox. What is a `firewall' for one observer, is totally transparent to the other.

Spontaneously breaking conformal symmetry is easy at first sight. It happens automatically in the Einstein-Hilbert action. One multiplies the metric tensor \(g_{\m\n}\) with the square of a scalar dilaton field \(\f(\vec x,t)\), which takes over the role of our field \(\W\). A curious feature of gravity is that the functional integral over the \(\f\) field has to be shifted to a complex contour, such that the vacuum value of the field becomes (see ref~\cite{GtH})
	\be\bra\f\ket = \pm i\sqrt{3\over 4\pi G}\ , \eel{phivac}
where the sign can be chosen freely.  The factor \(i\) in this equation is a notation that ensures that the \(\f\) field has the same unitarity and positivity properties as other scalar matter fields. Gauge-fixing the field to have exactly the value \eqn{phivac} reproduces standard Einstein-Hilbert gravity, and our description of the physical world will be as usual.

The power of our considerations comes if we decide te leave our \(\f\) field alone, and use \emph{something else} to fix the conformal gauge, a consideration that was absolutely crucial in understanding how the BEH-mechanism turns the electroweak theory into a manageable, that is, renormalizable, theory. Look at \eqn{phivac} as the ``unitarity gauge". We get a ``renormalizable gauge" if we decide to choose our conformal factor \(\W(\vec x,t)\) in such a way that the \emph{amount of activity} in a given space-time volume element is fixed or at least bounded. How to implement such a gauge choice is not known today.

Black holes still being black holes raises the question what happens to baryon number non conservation. Here, we point out that, even in the Standard Model, baryon number is not exactly conserved due to chiral anomalies.

The conspicuous conformal symmetry that seems to be only spontaneously broken in the Einstein-Hilbert action, is in fact also explicitly broken by anomalies. The conclusion we arrive at is, that the chiral anomaly is fine, but we can only accept matter interactions that are such that all scaling anomalies cancel out. This would be an extremely important constraint on the matter interactions as presently described in the Standard Model. This model will require drastic modifications at very high energies. The ensuing algebraic constraints may well lead to important clues as to the nature of our world at energies beyond the reach of particle accelerators, exactly as we hoped for when we turned our attention to the local conformal group.

{\footnotesize

}

\end{document}

How \emph{does} one measure the scalar component of \(T_{\m\n}\) in practice? We cannot use light rays to measure its effect on gravity. Normally, one would use light to measure the matter density, and use the equation of state to determine the pressure, but what if we do not know the equation of state, because we do not know the chemical composition of the material? Suppose that a local conformal transformation is possible that affects matter in a complicated way, so that its effect on gravity takes the form \eqn{Rtransf}? Of course, for atoms and molecules, this will not be easy, but I am thinking of a description of matter somewhere near the Planck scale, where we are forced to reconsider almost everything anyway. Then, the possibility of such transformations will affect physics in a drastic way, and this is what we should be interested in.